\newcommand{\A} {\hat{A}}
\newcommand{\As} {\hat{A}_s}

\newcommand{\Aout} {\hat{A}_s^{\mathrm out} }
\newcommand{\Aoutdag} {\hat{A}_s^{\dagger \, \mathrm out} }
\newcommand{\ah} {\hat{a}}
\newcommand{\as} {\hat{a}_s}

\newcommand{\ap} {\hat{a}_p}

\newcommand{\apc}{{\alpha}_p}

\newcommand{\rhop}{ \rho_p}  
\newcommand {\Gone } {G^{(1)} }
\newcommand {\chidue } {\chi^{(2)} }

\newcommand{\DD}{\mathcal D}                   
\newcommand{\D} { D_\textsc{pw}}                                      
\newcommand{\taugvm}{\tau_{\textsc{gvm}}  }
\newcommand{\lwoff}{\l_{\textsc{woff}}  }
\newcommand{\Deltapi}{{\Delta_p}_i}

\newcommand{\w}{\vec{w}}			

\newcommand{\vxi}{\vec{\xi}}	
	
\newcommand{\Om}{\Omega}
\newcommand{\q}{\vec{q}}

\newcommand{\rr}{\vec{r}}
\newcommand{\llangle} { \left \langle}   	
\newcommand{\rrangle} { \right\rangle}   	

\newcommand{\ex} {\vec{e}_x}
\newcommand{\ey} {\vec{e}_y}

\newcommand{\et} {\vec{e}_t}

\newcommand{\bsub}{\begin{subequations}}
\newcommand{\esub}{\end{subequations}}
\newcommand{\beq}{\begin{equation}}
\newcommand{\eeq}{\end{equation}}
\newcommand{\beqa}{\begin{align}}
\newcommand{\eeqa}{\end{align}}
\documentclass[fleqn,10pt]{wlscirep}
\usepackage[utf8]{inputenc}
\usepackage[T1]{fontenc}
\usepackage{amsmath} 
\usepackage{empheq}
\usepackage{amssymb}
\usepackage{graphicx}
\usepackage{float}   
\usepackage{braket}
\usepackage{xcolor}

\DeclareMathSymbol{\Rho}{\mathalpha}{operators}{"50}
\title{Modeling the space-time correlation of pulsed twin beams}
\author[1,*]{Alessandra Gatti}
\author[1,2]{Ottavia Jedrkiewicz}
\author[2]{Enrico Brambilla}
\affil[1]{ Istituto di Fotonica e Nanotecnologie del CNR, Piazza Leonardo da Vinci 32,  20133 Milano, Italy}
\affil[2]{Dipartimento di Scienze e Alta Tecnologia, Universit\`a dell'Insubria, Via Valleggio 11, 22100 Como, Italy}
\affil[*]{Alessandra.Gatti@ifn.cnr.it}
\begin{abstract}
Entangled twin-beams generated by  parametric down-conversion  are among the favorite sources for imaging-oriented applications, due their multimodal nature in space and time. However, a satisfactory theoretical description is still lacking. 
In this work we propose a semi-analytic model which aims to bridge the gap between time-consuming numerical simulations and the unrealistic plane-wave pump theory.  The model is used to study the quantum  correlation and  the coherence in the angle-frequency domain of the parametric emission, and demonstrates a  $g^{{1/2}} $ growth of their size  as the gain $g$ increases, with a  corresponding contraction of the space-time distribution. These predictions are systematically compared with the results of stochastic numerical simulations,  performed in the Wigner representation, of the full model equations: an excellent agreement is shown even for parameters well outside the expected limit of validity of the model. 
\end{abstract}
\begin{document}
\flushbottom
\maketitle
\thispagestyle{empty}
\section*{Introduction}
\begin{figure}[ht]
\centering
{\scalebox{.68}{\includegraphics*{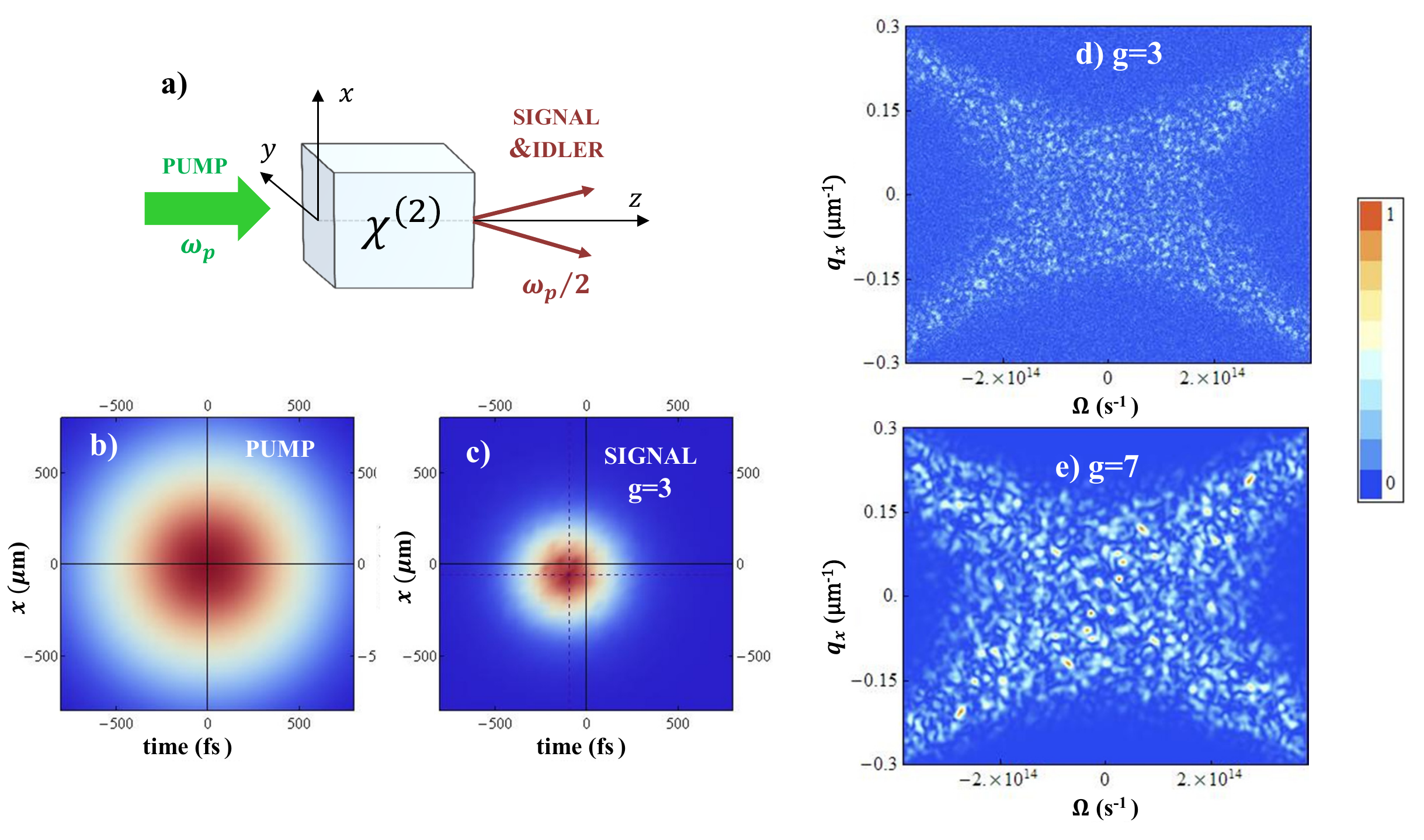}}}
\caption{a)  Scheme of the PDC process. (b) and (c) Mean space-time distributions  of the pump and of the down-converted signal, respectively, at the crystal output face. d) and e) Instantaneous signal distribution in the frequency and transverse wave-vector domain, exhibiting  a  broad X-shaped spectrum with  small and correlated  speckles, whose  size  increases with the gain $g$. Data  taken from quantum simulations of a 2 mm BBO  (Beta-Barium Borate) crystal, cut  for collinear phase-matching at $\lambda_p=515 $nm, pumped by  a Gaussian beam of waist $w_p=600 \, \mu$m and duration $\tau_p= 600$ fs.  }
\label{fig_Xspectrum}
\end{figure}

Modern nonlinear optics not only  enabled  the first fundamental tests of quantum mechanics, but  also paved the way for the advent of  quantum technologies.  In this context, a central role  has been played  by parametric down-conversion (PDC), a process in which photons of  a laser   beam that propagates 
 in a nonlinear $\chidue$ crystal occasionally split into photon pairs  at lower energy (Fig \ref{fig_Xspectrum}a).  The microscopic mechanism of pair creations    is at the origin of a {\em high dimensional entanglement }  \cite{Law2000,Eberly2004,Gatti2012,Malte2009}, both in the sense that paired photons are quantum correlated in different degrees of freedom (polarization, transverse position-momentum, time-frequency)  and that the entanglement involves a huge number of independent modes due to the ultrabroad bandwidths of PDC. These features, which  persist also  at the  macroscopic level of  bright entangled  beams,  made  PDC a favorite choice  for quantum imaging applications (see the reviews  \cite{Gatti2008,Genovese2016,Padgett2019}): 
 the transverse entanglement of photon pairs was indeed used for  pioneering realizations of ghost imaging \cite{Pittman1995} and for
 quantum enhanced  microscopy \cite{Ono2013}, while  the  sub shot-noise spatial correlation of twin beams 
 allowed to demonstrate high-sensitivity   imaging  \cite{Brambilla2008,Brida2010}, just to cite few examples.  The time-frequency entanglement is  at the basis of the  quantum enhancement of two-photon absorption,  used in entangled two-photon microscopy \cite{Tabakaev2021,Raymer2021b} and spectroscopy \cite{Schlawin2018}. 
\par 
Despite  its extensive exploitation, the theoretical description of  multimode  PDC is still unsatisfactory.  A complete analytical model  exists\cite{Klyshko1988} 
only  for the  low-gain regime of spontaneous PDC (SPDC), and it is widely-used to describe the space-time quantum correlation of photon pairs \cite{Law2000,Eberly2004, Gatti2012}. On the other side,  the bright twin   beams generated by high-gain PDC are becoming, for obvious reasons,  more and more attractive for imaging oriented applications. However,  the description of their spatio-temporal correlation  needs  to  resort   to time-consuming numerical simulations \cite{Brambilla2004} or  to the nonphysical limit of a plane-wave  pump \cite{Klyshko1988,Brambilla2001,Gatti2003}.
\par
Purpose of this work is to present a simple semi-analytic  model for PDC that accounts for  the finite duration and transverse cross-section of a pulsed pump in  any gain regime. The model will be derived on the basis of heuristic arguments, and its predictions systematically compared with the results of numerical simulations,  performed in the quantum domain, of the full model equations. 
We shall focus on those aspects which are not encompassed by the plane-wave pump (PWP) theory, namely:  
\begin{itemize} 
\item The  correlation and coherence volumes of PDC light  in  the angle-frequency domain, which can be observed in the far-field of the source. 
 A generally accepted view is that it is sufficient to substitute the Dirac-delta correlations of the PWP theory with  finite peaks  given by the angle-frequency spectrum of the pump laser, as predicted in the spontaneous regime.  However, various experiments \cite{Jedr2006, Allevi2014}
observed a substantial increase of the coherence and correlation lengths    with increasing gain (see also the examples from numerical simulations in  Fig.\ref{fig_Xspectrum}d,e). 
\item The space-time distribution of PDC light,  in the near-field of the source. Again, the intuitive view is that  it should superimpose  to the pump pulse as it happens in the spontaneous regime, but, as we shall see, an important shrinking takes place at high or even moderate gain (Fig.\ref{fig_Xspectrum}c). 
\end{itemize} 
The formulation of this model is motivated  by both practical and fundamental reasons.  Today's implementations of high gain PDC use short laser pulses (hundreds of femtoseconds ), which  
 by no means  fit  into  the PWP description. Moreover,  it is not clear whether chirping the pulse, as commonly done in nonlinear optics, is  suitable for quantum applications.   
On a more fundamental side,  contrary to the SPDC regime  \cite{Law2000,Eberly2004, Malte2009,Gatti2012},  a quantitative assessment of  the global entanglement of  multimode high-gain PDC   is currently not easy,  because the PWP model is clearly  unable  to provide  the number of entangled modes.  This is of paramount importance in many ambits,  as e.g.  in order  to evaluate the quantum enhancement  of   two-photon absorption\cite{Raymer2021b}. 
\section*{Theory Background}
Let us consider  two wavepackets associated with the pump field (central frequency  $\omega_p$), and with the downconverted signal   ($\omega_s={\omega_p}/{2}$) that propagate inside a nonlinear $\chidue$ crystal of length $l_c$ forming small angles with a mean direction $z$.  Let 
\begin{align}
\A_j(\rr, t,z)   &=  \int \frac{d^2 \q}{2\pi}   \int \frac{d\Om}{\sqrt{2\pi}}  e^{i \q \cdot \rr } e^{ - i \Om  t }   \A_j (\q, \Om, z)   \qquad j=s,p 
\label{AFourier}
\end{align}
be their electromagnetic field operators, where  $\rr=x\ex +y \ey$  is the  position in the transverse plane,  $\q= q_x \ex + q_y \ey $  is the transverse  wave-vector,   $\Om$ is the  frequency offset from the carriers,    and   dimensions are such that 
 $ \A_j^\dagger (\rr,t,z)  \A_j  (\rr,t,z)$  is  a photon number per unit area and time.  
Their evolution along the slab is best described in an interaction picture in which the fast linear propagation is subtracted  (see \cite{Gatti2003} for details), by introducing 
\beq
\hat a_j (\q,\Om,z) = e^{- i k_{z j} (\q,\Om) z}  \A_j (\w,z) ,  \qquad \text{where} \quad k_{jz} (\q, \Om) = \sqrt {k_j^2(\q,\Om ) -q^2}  \, , 
\label{ajdef}
\eeq
$ k_j(\q, \Om)$ 
being  the wave-number of the j-th wave (it  depends on the direction of propagation through $\q$ only for the extraordinary wave). While simulations will consider the more general coupled equations \eqref{NLeq},  in the analytics we shall exploit the undepleted pump approximation in which  the  pump operator  $\ap (\q,\Om, z) \approx \ap (\q,\Om, 0)  $  is substituted by a c-number field.   By adopting a shorthand notation,  
 in which  $\vxi:  = (\rr, t) \in R^3   $   is the  space-time  vector and  $\w : = (\q, \Om)  $ is  its conjugate Fourier vector, with the convention for the scalar product: $ \w\cdot \vxi := \q\cdot \rr  - \Om t $, the evolution of the PDC field along the slab is  then described by: 
\begin{align} 
\frac{\partial  \as }{\partial z}   (\w, z )   &=  \frac{g}{l_c}  \int 
 \frac{d^3 \w_0 }  {(2\pi)^{\frac{3}{2}} }  \, \apc (\w_0 ) \, 
 \as^\dagger(\w_0-\w, z)  e^{-i \DD (\w;  \w_0-\w) z }  ,
\label{prop}
\end{align}
where:  $g$ is the  dimensionless gain parameter, proportional to the nonlinear susceptibility, the crystal length and the pump peak amplitude; 
$\apc (\w)$ is the Fourier profile of the input pump field, normalized so that $\apc (\vxi =0)=1$; 
\begin{align}
\DD ( \w;   \w_0 - \w)  &:=  k_{sz} (\w) +  k_{sz} (\w_0 - \w )  - k_{pz} (\w_0) 
\label{PM}
\end{align}
is the phase mismatch of  a   down-conversion process  in which a  pump photon in mode $\w_0$  disappears and  a photon pair  is created  in modes $\w$ and $\w_0 -\w$, with conservation of the  energy  and transverse momentum. 
 The solution  of equation \eqref{prop} is a  generalized Bogoliubov-type transformation, similar to that studied in \cite{Brambilla2004}, linking in a nonlocal way the field operators at the crystal output $  \Aout (\w) $ to the input vacuum fields.  
We shall not deal here with  such a general solution, but rather focus on the two second order moments 
\begin{align}
& \Psi (\w_1, \w_2)  = \left\langle \Aout (\w_1) \Aout ( \w_2)  \right\rangle 
\label{Psi}  \\
& \Gone (\w_1, \w_2) =\left\langle \Aoutdag (\w_1) \Aout \w_2)  \right\rangle  
\label{G1}
\end{align}
that for such a Gaussian process  determine all the spatio-temporal statistical properties at the medium output.   
The first function,  is the  {\em biphoton correlation}, proportional to   the probability amplitude of generating a pair of  twin photons  in modes $\w_1$ and $\w_2$.
The second one 
 is  the ubiquitous coherence function,  describing   the autocorrelation of light when twin photons are not detected together, for example, by measuring only one side of the spectrum with respect to the central frequency. 
All higher order moments can be expressed in terms of the second order ones: for example the correlation  of  the light intensity $\hat I (\w) = \Aoutdag (\w) \Aout (\w) $ reads: 
$\langle : \hat I (\w_1)  \hat I (\w_2) :\rangle  - \langle \hat I (\w_1) \rangle  \langle \hat I (\w_2) \rangle =
| \Gone (\w_1, \w_2) |^2 + \left| \Psi (\w_1, \w_2)  \right|^2 $. 
\par
 Goal of the next sections will be to determine the functional form of  $\Psi $ and $\Gone$, holding under specific approximations. Before that, let us remind the well-known  results of the plane-wave pump model \cite{Klyshko1988}, in which  $\apc (\w) \to (2 \pi)^{3/2} \delta (\w)$, and 
\begin{align}
& \Psi (\w_1, \w_2)  \stackrel { PWP} { \to } \delta (\w_1 + w_2) \, U(\w_1) V(-\w_1)  e^{ik_p l_c} \qquad 
& \Gone (\w_1, \w_2)   \stackrel { PWP} { \to } \delta (\w_2 - \w_1) \, \left| V (\w_1)\right|^2 \, . 
\label{PWcorr}
\end{align} 
where 
$ 
U(\w) =
\cosh{ \left[\Gamma (\w) \right]  } + i\frac{\D (\w)l_c}{ 2 \Gamma (\w)} 
\sinh \left[\Gamma (\w) \right] 
$ ;     $ V(\w) = 
 g  \frac{\sinh \left[\Gamma (\w) \right]}{ \Gamma (\w)} 
$, with 
$\Gamma (\w)= \sqrt{g^2 - \frac{[\D (\w)l_c]^2}{4} }  $. The two functions are  strongly peaked in the regions where the plane-wave phase mismatch
\beq
\D (\w) := \DD (\w, -\w) 
\eeq
vanishes. In the example of Fig.\ref{fig_Xspectrum}, where parameters are chosen for collinear phase matching, it corresponds to the broad X-shape  of the  Fourier spectrum.  The PWP theory has the undoubted merit of describing very well the  angle-frequency distribution of PDC light  in any gain condition  \cite{Jedr2006a,Gatti2009,Caspani2010,Jedr2012b}, and the existence of non-classical correlations between  conjugated Fourier modes (see e.g. \cite{Kolobov1999,Brambilla2001,Gatti2008}). An obvious drawback  are  the Dirac-delta correlations in the Fourier domain, direct consequence of assuming  a homogeneous distribution in the space-time domain.
\section*{Results}
\subsection*{Our proposal: the quasi-stationary  model}
With the aim of formulating a model valid both in the low and high gain of PDC, but without the heavy limitations of the plane-wave pump approximation, we focus on  a sufficiently narrowband pump, such that 
 the pump Fourier profile   dies out on a faster scale than the phase matching function: 
\beq
\DD(\w,\w_0-\w)  l_c \simeq \D (\w)  l_c  + \Om_0 \taugvm  + q_{0x}  \lwoff  \to \D(w) l_c  \qquad  \forall  \, \w_0 \text{ inside the pump spectrum}. 
\label{NPWP}
\eeq
where the Taylor expansion \eqref{Taylor} has been used.  
Here $ \taugvm=\frac{ l_c} {v_{gs}} -\frac{ l_c}{v_{gp}}$ and   $\lwoff =- l_c  \rho_p$  are the overall temporal and spatial walk-off  occurring between the  signal and  the pump during propagation  (Methods), due to their group velocity mismatch and  the walk-off of the Poynting vector, respectively.  
Strictly speaking  the limit \eqref{NPWP} requires a pulse  of duration $\tau_p \gg \taugvm$ and waist $w_p \gg \lwoff$:  in 
 the spontaneous regime of PDC, these   conditions  ensure that  both the biphoton  correlation   and the coherence function factorize into the product of two functions with  different scales of variation\cite{Gatti2009, Caspani2010}. 
In  high-gain  we do not have such  explicit expressions. Our approach  
 will be to assume that  in any regime  the $\Psi$ and $\Gone$  factorize in the product of a fast decaying correlation peak and a slowly varying envelope. 
This ansatz, to which we shall refer as {\em Quasi-Stationary}  (QS) approximation,  corresponds to a configuration where small speckles are observed inside a broader spectral distribution, as depicted by the simulation of Fig.\ref{fig_Xspectrum} and  observed in several high-gain experiments \cite{Jedr2004,Jedr2006a,Jedr2012b,Allevi2014} We expect that such separation of decay scales  holds in the  limit \eqref{NPWP}, but we shall verify this point a posteriori. As for the specific choices of envelopes and correlation,  we set: 
\begin{subequations}
\label{QSmodel}
\beq
\begin{aligned} 
&\Psi (\w_1, \w_2)   =   \mu_{corr} (\w_1+\w_2) \,  U(\w_1) V(\w_1)  e^{i k_p l_c} \, , \;  \text {with}  
\label{PsiQS} \\
&\mu_{corr} (\w )    =    \int \frac{d^3 \xi}{(2\pi)^3}   e^{-i  \w  \cdot \vxi}     \, \frac{  \sinh {[ 2 g |\apc (\vxi -\vxi_M)| ]}    }{ \sinh{2g}}   e^{i \phi_p (\vxi -\vxi_M)}   
\end{aligned}
\eeq
where $ \phi_p  (\vxi)= \arg{[ \apc  (\vxi)]} $, and $U$ and $V$ are  the functions of  the PWP model  \eqref{PWcorr}, and 
\beq 
\begin{aligned} 
&\Gone (\w_1, \w_2)     =   \mu_{coh} (\w_2-\w_1) \,  \left| V(\w_1) \right|^2 \;  \text {with}  
\label{G1QS}\\
& \mu_{coh}  (\w)    =    \int \frac{d^3 \xi}{(2\pi)^3}   e^{-i  \w \cdot \vxi}  \,   \frac{    \sinh^2 {[g |\apc (\vxi - \vxi_M)| ]} }{ \sinh^2{g}}  
\end{aligned}
\eeq
\end{subequations}
 In these  formulas 
$\rr=0$ and $t=0$ are the coordinates of the pump center at the crystal { output}. Accordingly,  
\beq 
\vxi_M= \frac{\lwoff}{2}   \ex   + \frac{\taugvm}{2}  \et
\label{XiM}
\eeq
 represents, as we shall see,   an offset (in space and time)  between  the signal and pump beams at the crystal output face. 
Clearly, in the limit of a  homogeneous and stationary pump  in which  $\apc(\vxi)=1$, 
the  PWP results of Eq.  \eqref{PWcorr} are recovered.  On the other hand,  the functions $\mu_{corr} $ and $\mu_{coh}$ that replace the Dirac-delta of the PWP model
 have  been chosen  on the basis of  an other  notable limit, namely that of a {\em thin crystal}, in which one takes  $\D(\w) l_c \approx 0$  for all the modes of interest. Then,  the propagation equation \eqref{prop} can be easily solved (Methods), and the Eqs.\eqref{QSmodel} converge asymptotically to the results thereby obtained. 
Notice that the { thin crystal} limit, which  can be  also interpreted as discarding the contribution of modes which are not phase-matched, is  complementary to the PWP limit, because it assumes a homogeneous distribution in the Fourier domain. 
\subsection*{Correlation and coherence volumes in the Fourier space}
The  QS model of Eqs.\eqref{QSmodel}  merges the "good" features of  the plane-wave pump and thin crystal approximations, by giving asymptotically the same results, but being free from their ill behaviors.
Said that, then a good question is whether  this model makes correct predictions in realistic conditions, and what are their limits of validity. To this end,  we shall compare the predictions of the QS model with the results of  numerical  stochastic simulations of the complete model \eqref{NLeq}.  
\par
Let us focus on a  real and symmetric pump , i.e. a pump  that is not chirped and  that satisfies  $\apc(-\vxi) =\apc(\vxi)$ . Then, 
the widths of the spectral correlations \eqref{PsiQS} and \eqref{G1QS} can be straightforwardly evaluated.  
 By defining 
\beq
\begin{aligned}
&F_{ corr} (\vxi) =\frac{   \sinh {[ 2 g |\apc (\vxi)| ]}      } { \sinh{2g}}, \qquad \qquad 
   & F_{ coh} (\vxi) = \frac{  \sinh^2 {[  g |\apc (\vxi)| ]}    }{ \sinh^2{g}} 
\end{aligned}
\label{Fbeta}
\eeq
we have 
\begin{align} 
&\left| \mu_\beta (\w)  \right|=  \int \frac{ d^3 \xi}{(2 \pi)^3}   e^{-i \vxi \cdot \w}  \, F_\beta (\vxi) ;   \qquad
 &   \int d^3 w   \, \left| \mu_\beta (\w)  \right|  
= \int d^3 \xi  F_\beta (\vxi)  \, \delta (\vxi) = F_\beta (0) =1  \qquad (\beta=corr,\, coh )
\end{align}
Hence, each $ \left| \mu_\beta \right|  $  defines a  normalized distributions,  having $ F_{\beta} $ as its characteristic function. Their mean values  vanish for symmetry reasons,   which gives the usual results that the peak of the coherence is at   $\w_2= \w_1$, while  the biphoton correlation is peaked at $\w_2= -\w_1$. Their widths  can be evaluated as 
\beq
\begin{aligned}
\langle   w_i^2  \rangle_\beta  -\langle   w_i  \rangle_\beta^2   &= \int d^3 w \left |\mu_{\beta} (\w) \right|  \, w_i^2
=  \int d^3 w \int \frac{d^3 \xi} {(2\pi)^3} \left[  -\frac{\partial^2}{\partial \xi_i^2}  e^{-i \vxi \cdot \w}\right] F_\beta (\vxi) 
&= - \left. \frac{ \partial^2 F_\beta (\vxi)}{\partial \xi_i^2}    \right|_{\vxi =0}  
\end{aligned} 
\eeq
where $w_i $ and $\xi_i \, $  ($i=1,2,3$) denote  the components of the vectors $\w$ and $\vxi$, and 
integration by parts has been performed twice.    
By focusing on a   a Gaussian pump of the form 
$
\apc (\vxi) = \exp{ ( -\sum_i \frac{\xi_i^2}{ \Deltapi^2} )}
$
  where $\Deltapi= w_{px}, w_{py}, \tau_p $ are the 1/e widths in the  spatial  and temporal directions, we have 
$
\langle  w_i^2 \rangle_{corr}    
=   \frac{2g} {\tanh (2g)} \frac{2}{\Deltapi^2}  $, and 
$ 
\langle  w_i^2 \rangle_{coh}    
 =   \frac{2g} {\tanh (g)} \frac{2}{\Deltapi^2}
$. 
The QS model provides in this way  simple and explicit  formulas for the correlation and coherence lengths in the Fourier space, defined here 
as the standard deviations  of the respective distributions $|\mu_{\beta}|$
\bsub
\label{varcorr} 
\begin{align}
&\Delta w_{i \, corr}   = \sqrt{\langle  w_i^2 \rangle_{corr}    }  
= \sqrt{ \frac{4g} {\tanh (2 g)}  } \frac{1}{\Deltapi}
\label{varpsi}
\\
& \Delta {w_i}_{ coh}  = \sqrt{\langle w_i^2 \rangle_{coh}    } 
= \sqrt{ \frac{4g} {\tanh (g)}  }  \frac{1}{\Deltapi}
\end{align}
\label{varG1}
\esub
\begin{figure} [ht]
{\scalebox{.67}{\includegraphics*{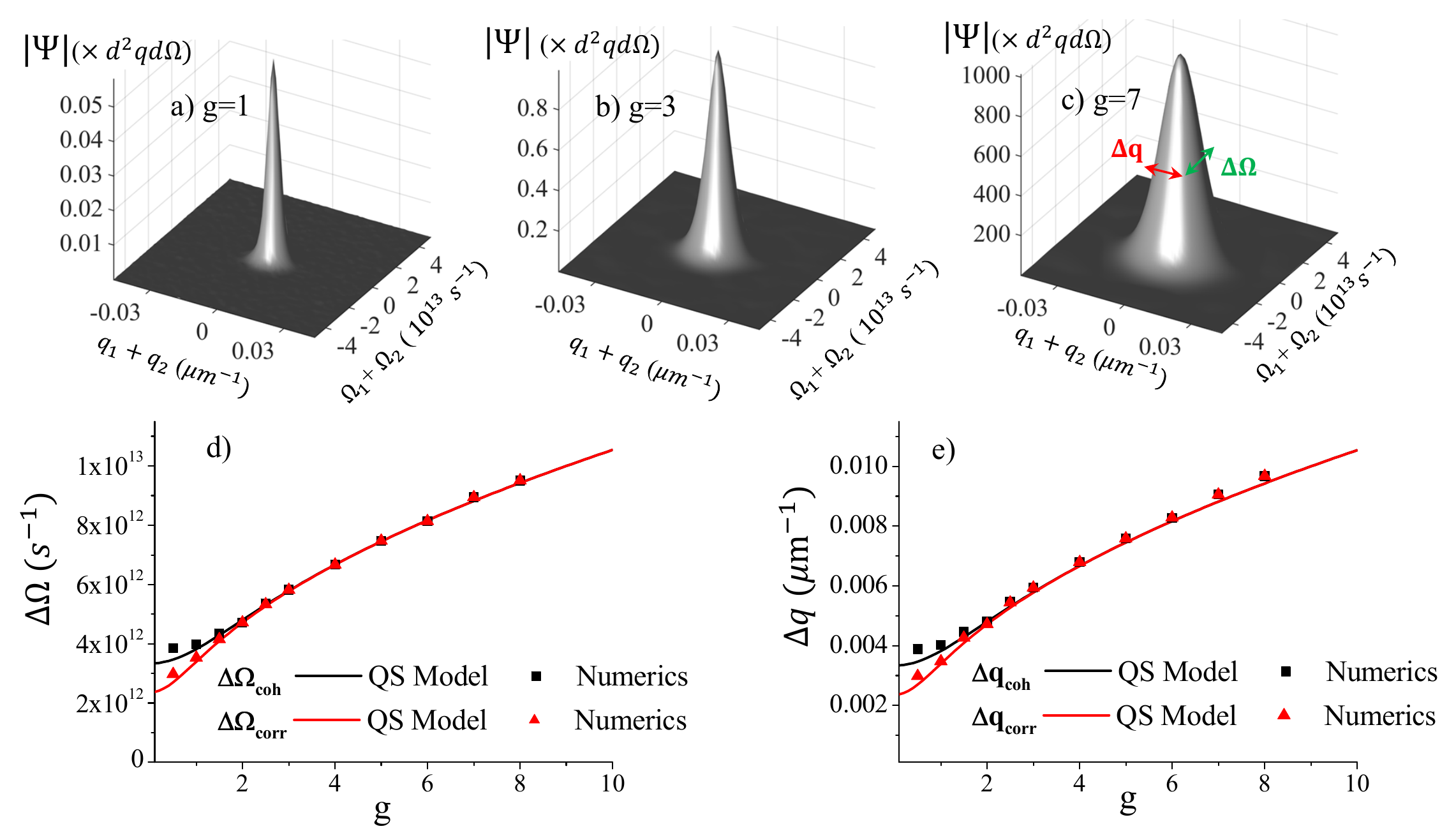}}}
\caption{"Long" pump $\tau_p= 600$ fs, $w_p=600 \, \mu$m. 
a) , b),  c): Examples of the biphoton correlation ($\left| \Psi\right|$, multiplied by the volume $d^2q d\Om$ of the simulation pixel),  numerically calculated by averaging over 3000-10000 stochastic realizations of Eqs.\eqref{NLeq},  showing the increase of the correlation volume with gain. 
d)    Spectral width (standard deviation) $\Delta \Omega$  of $\left| \Psi \right|$  (red) and of 
 $| \Gone |$  (black) as a function of the gain $g$. 
Solid lines: predictions of the QS model, according to Eqs.\eqref{varcorr}. Symbols: results of simulations. e) Same  as d), but for the width along $q$. } 
\label{fig_corr600}
\end {figure} 
The two curves are plotted  by the solid lines in Fig.\ref{fig_corr600}.
For large gains,  the correlation and the coherence  functions have the same width, which increases with the gain  as $\Delta w_{i \, \beta} \sim   \sqrt{4g}/\Deltapi $.   
 This is in nice agreement with the experimental findings \cite{Allevi2014}, 
where a $g^{1/2}$ growth of the size of the speckles was observed  in the angle-frequency domain  of  high-gain PDC. 
At small gains,  the two curves separate,
with $ \Delta w_{i\, corr}  \to \sqrt{2} /\Deltapi $, and  $ \Delta w_{i\, coh}  \to 2/\Deltapi $, which are just the inverse of the standard deviations  of the pump amplitude and intensity, respectively. 
Actually,   we notice that  in  the limit $g\ll 1 \, $  
$\mu_{corr} (\w) $ becomes the Fourier transform of the pump {\em amplitude}, while 
 $\mu_{coh} (\w) $ becomes the Fourier transform of the pump {\em intensity}, in agreement with the results known  for spontaneous  PDC\cite{Gatti2009,Caspani2010}. 
\begin{figure} [ht]
{\scalebox{.67}{\includegraphics*{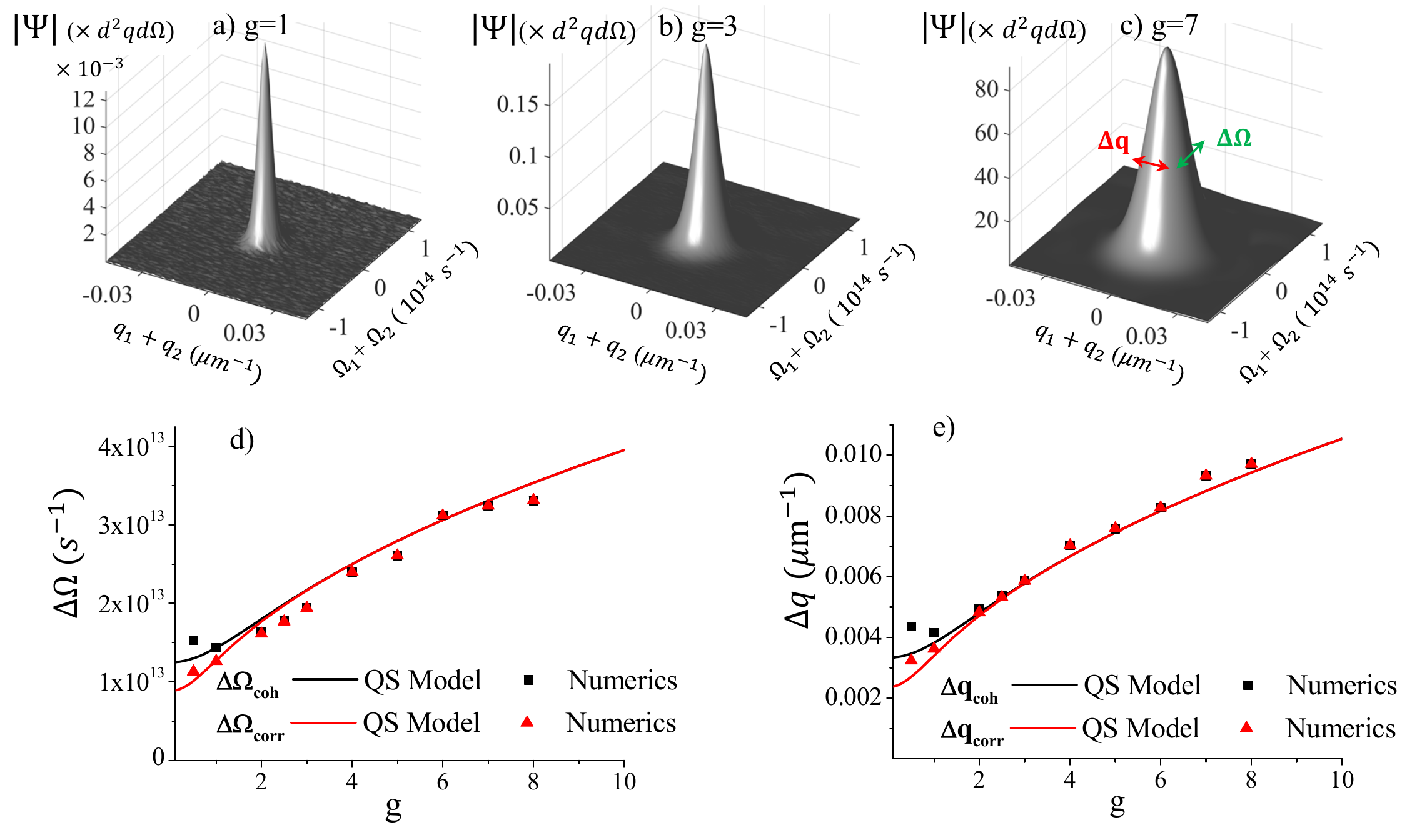}}}
\caption{"Short" pump $\tau_p= 160$ fs, $w_p=600 \, \mu$m. Same as Fig.\ref{fig_corr160}, except that  up to $10^5$ stochastic  realizations were necessary to obtain the numerical results, due to the high noise. 
Notice the  fair agreement with the  QS model predictions, despite $\tau_p < |\taugvm|$.} 
\label{fig_corr160}
\end {figure} 
\\
Superimposed to the analytical predictions, the symbols show for comparison the results of the complete model, obtained by  stochastic 2D+1 simulations of  the propagation  equations \eqref{NLeq},  performed in the Wigner representation (Methods). We considered   a 2mm BBO crystal cut for degenerate and collinear phase matching  at $\lambda_p=515\,$nm;   in these conditions, the group velocity delay is $\taugvm=- 185.6\,$fs, while  the lateral walk-off is $\lwoff=-113 \, \mu$m.  The pump duration  and waist  ($\tau_p=600 $ fs and  $w_p = 600 \mu$m, respectively)  were chosen reasonably larger than these values,  in order to meet  the condition \eqref{NPWP} of a  {nearly plane-wave} pump.  Indeed,  for these parameters the predictions of our simple QS model appear in excellent agreement with   the results of simulations. 
Conversely,  figure \ref{fig_corr160} shows the results of  simulation of  PDC from a much shorter pulse  $\tau_p=160\,$fs,  which  is well outside the expected conditions of validity of the QS model. Nevertheless, the agreement with the QS model is still fairly satisfactory: numerical results are more scattered than in Fig.\ref{fig_corr600}, but  follow nicely the behaviors  of  a $\sqrt{4g} $ growth at high gain, and a bifurcation at low gain.  At low gain the agreement is not perfect, but this may also  be due to the impact of residual noise. 
%
\subsection*{Exponential narrowing   of the space-time distributions}
\begin{figure} [ht]
\centerline{\scalebox{.6}{\includegraphics*{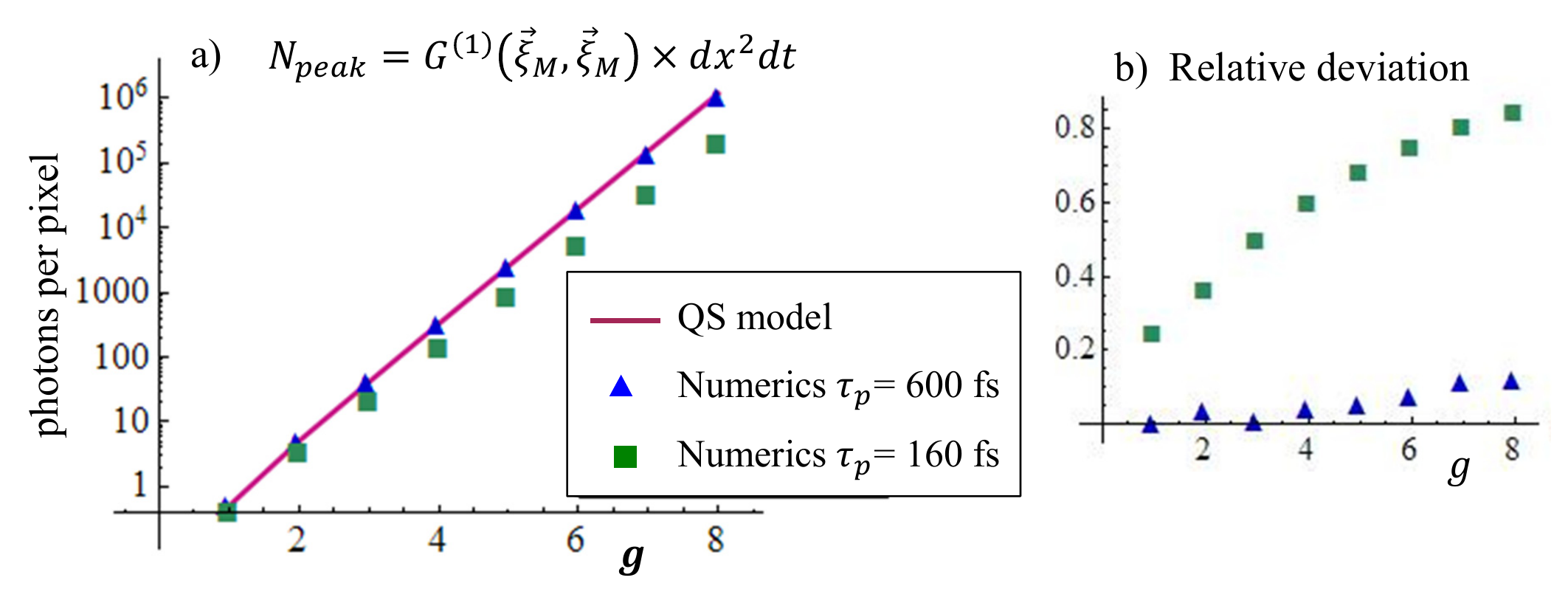}}}
\caption{  a) Mean number of photons at the peak of the signal pulse  $ \langle \Aoutdag (\vxi_M) \Aout (\vxi_M) \rangle \times dx^2 dt$ (units are photons per numerical pixel). The red line is the QS model prediction  in Eq. \eqref{Is}, where the integral was calculated via a discrete sum over the same  numerical grid of the simulations. The symbols are the results of simulations. b) Relative deviation between analytic predictions and numerical results}
\label{fig_Npeak}
\end {figure}

The growth   of the correlation and coherence volumes  in the Fourier domain  with increasing gain is strictly linked  to 
 a  progressive  narrowing of the space-time distribution of PDC photons at the medium output:   at low gain these simply follow the profile of the pump, because at each point of the medium   the  photon  pairs are generated independently  with probability proportional to  $|\alpha_p (\rr,t)|^2$.  At high gain the signal distribution  becomes  much narrower, because  of  stimulated processes that exponentially  grow in the central region of the pump peak.  
More formally, by Fourier transforming  the spectral correlations in Eqs.\eqref{QSmodel}, one obtains
\bsub
\begin{align}
& \Psi (\vxi_1, \vxi_2)  = \llangle \Aout(\vxi_1) \Aout (\vxi_2)  \rrangle =
F_{corr} (\vxi_1-\vxi_M) \int \frac{d^3 w} {  (2 \pi)^3} \,  e^{i (\vxi_2 -\vxi_1) \cdot \w} U(\w) V(\w)  \\
& \Gone (\vxi_1, \vxi_2) = \llangle \Aoutdag (\vxi_1) \Aout (\vxi_2)  \rrangle =
F_{coh} (\vxi_1-\vxi_M) \int \frac{d^3 w} {  (2 \pi)^3} \,  e^{i (\vxi_2 -\vxi_1) \cdot \w} | V(\w)|^2
\end{align} 
\esub 
The Fourier integrals  at r.h.s. define   spatio-temporal correlation peaks  centered around $\vxi_2=\vxi_1$, known in the literature 
as  X-entanglement\cite{Gatti2009, Caspani2010,  Jedr2012b} and  X-coherence \cite{Jedr2006a}, because of their particular shape. We shall not deal here with these aspects, already extensively studied, but focus on the envelopes $F_\beta$, which,  at contrary  the PWP model, used e.g. in \cite{Caspani2010},  have  well defined expressions also at high-gain. 
 In particular,  the QS model provides the  mean spatio-temporal distribution of the PDC  photons in any gain regime:  
\beq
\llangle  \Aoutdag (\vxi) \Aout (\vxi)  \rrangle = \Gone (\vxi,\vxi)=  \frac{\sinh^2  [ g |\apc (\vxi-\vxi_M)|]  }{\sinh^2 (g)}
\int \frac{d^3 w} {  (2 \pi)^3}   | V(\w)|^2
\label{Is}
\eeq
\begin{figure} [ht]
\centerline{\scalebox{.7}{\includegraphics*{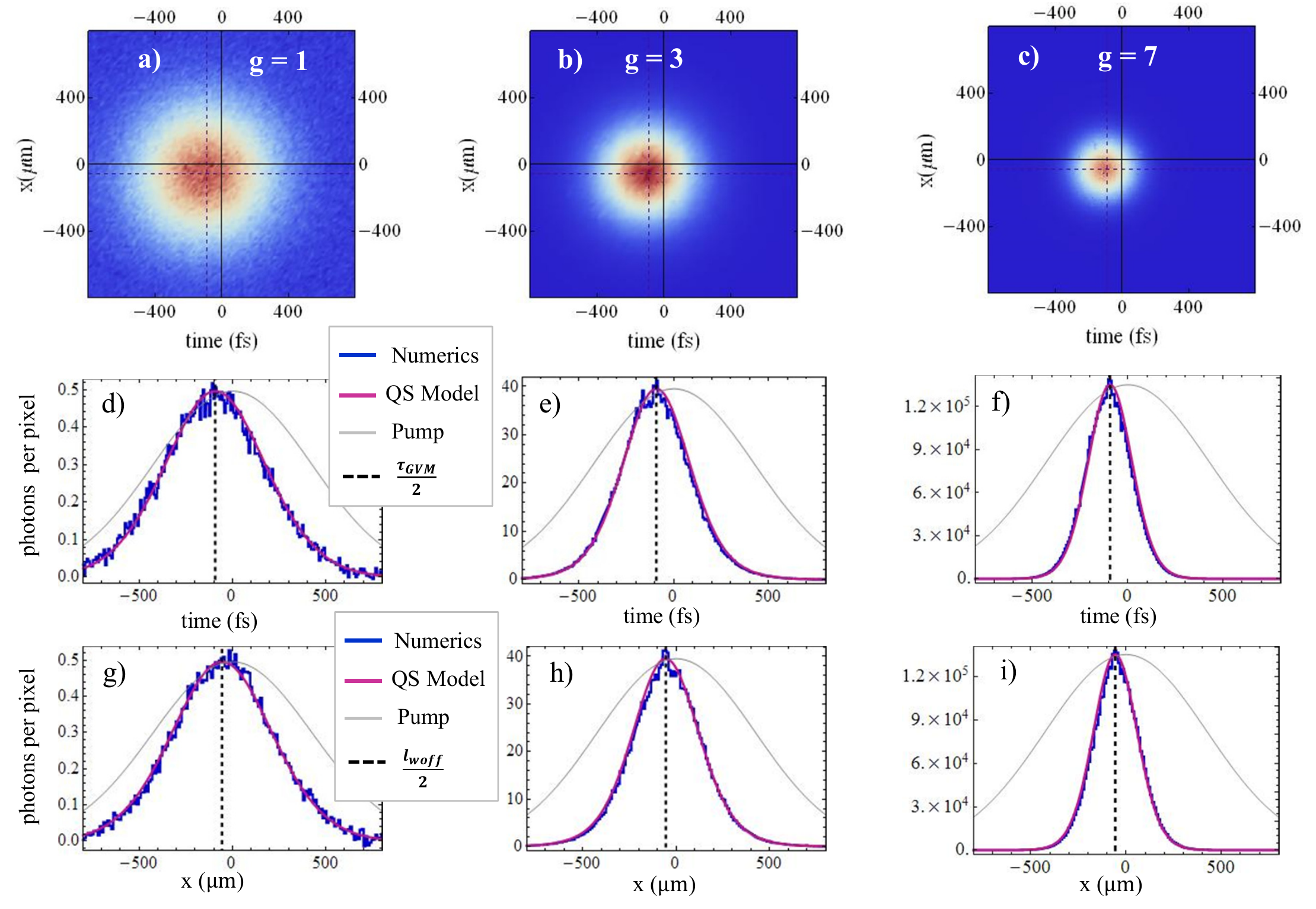}}}
\caption{ Space-time distributions  (mean photon number  on the pixel of the numerical grid)  of the PDC light at the crystal output, evidencing the narrowing of the signal beam with increasing gain. 
 "Long" pump $\tau_p= 600$ fs, $w_p=600 \, \mu$m. First row: 2-D plots from simulations. Second and third row: sections along time and space, respectively.  Superimposed to the numerics (blue), the red curves show the predictions \eqref{Is} of the QS model, multiplied by a fit parameter,  while the gray lines show the pump profile. $g=1$ in a), d)  and g). $g=3$ in b), e) and h). $g=7$ in c), f) and i). }
\label{fig_NF600}
\end {figure} 
The behavior of this function is  shown by the red lines in Figs.\ref{fig_NF600} and \ref{fig_NF160}. 
The QS model predicts that the  signal pulse is significantly narrower and shorter than the pump pulse (gray lines) even at moderate  gain as $g=1$; in addition, it  predicts that  it appears  at the  medium output  delayed  (actually anticipated) by a an amount $\frac{\taugvm}{2}$ and laterally  shifted by  $\frac{\lwoff}{2}$. The blue lines are instead the results of quantum simulations of the complete model, which  fully incorporates the effects of  temporal and spatial walk-off as well as of dispersion and diffraction. As already observed for Fig.\ref{fig_corr600}, the  QS model, despite its highly simplified formulation, provides an excellent description of the process. 
\begin{figure} [ht]
\centerline{\scalebox{.7}{\includegraphics*{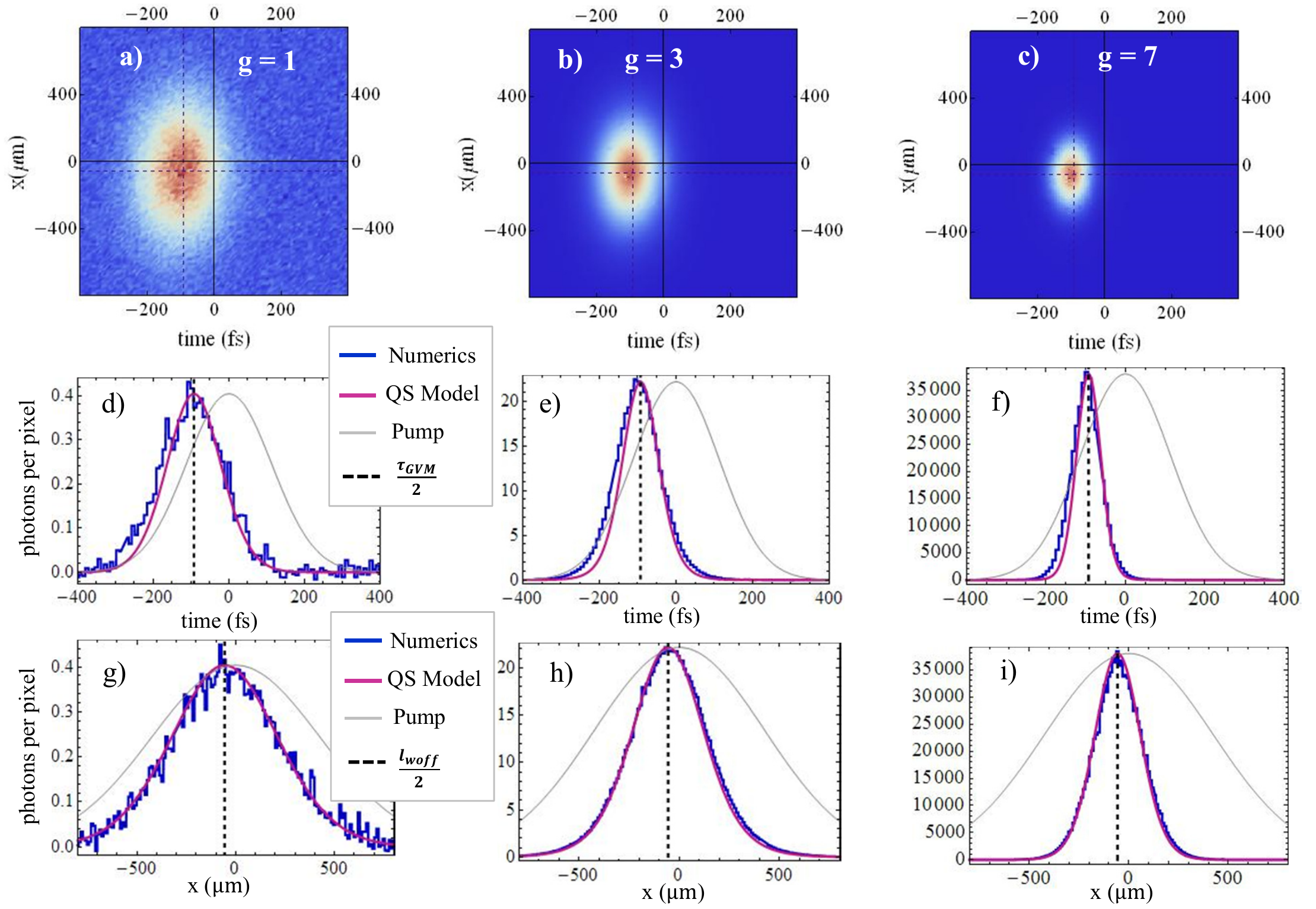}}}
\caption{ Same as figure \ref{fig_NF600}, but for a "short"  pump pulse $\tau_p= 160$ fs, $w_p=600 \, \mu$m.
 }
\label{fig_NF160}
\end {figure} 
The fit between analytics and numerics is almost perfect for the long pump $\tau_p=600 $fs in Fig, \ref{fig_NF600}, but remains very satisfactory also for the short pump  $\tau_p=160$fs  in Fig, \ref{fig_NF160}.  \\
In this regard, however, it should be noted that the graphs were produced with the aid of a fit parameter that multiplies the analytic function \eqref{Is}. This parameter was close to unity for the long pulse, but significantly smaller for the short one. In fact, as illustrated by Fig.\ref{fig_Npeak}, the QS model is able to provide a reasonable approximation of the mean number of generated photons, within a  $ 10 \% $ error, only in the case of the long pump pulse. On the contrary, it essentially fails in the case of a short pulse for which $  \tau_p \precsim  \taugvm$. This is predictable, as the QS model takes into account the effects of the spatial and temporal walk-off between the pump and signal only through a rigid traslation of the two distributions: as is known, however, the loss of overlap between the two beams during  propagation, due to their different group velocities and the spatial walk-off of the extraordinary pump, reduces the efficiency of the stimulated conversion processes.
Nevertheless, we note that even in the case of $ \tau_p = 160 $fs   where the  group velocity delay $\taugvm=-185$ fs  causes an important loss of overlap between the signal and pump 
 (see Fig. \ref{fig_NF160}d, e) and f)),   the QS model still provides a very  efficient description of the size and shape of the signal beam. 
\section*{Discussion}
We formulated a semi-analytical model describing  pulsed PDC in any gain regime and verified its predictions through simulations of the full model equations. The rationale behind the formulation of our model is i) the simplicity and accessibility of the results,  an example being the explicit formulas \eqref{varcorr}  for the correlation  and coherence volumes, and ii) that the model converge asymptotically to the appropriate limits in the cases where the solution of the evolution equation \eqref{prop}  is known (i.e. the PWP, the thin crystal and the $ g \to  0$  limits). 
The model essentially neglects the effects of the spatial and temporal walk-off  between the pump and signal,  except for a rigid translation  of the two distributions: as such we expect it to be valid only for pump pulses substantially longer than $\taugvm$ and broader than $\lwoff$. 
Surprisingly, we verified that its predictions regarding the shape and size of  the Fourier coherence and correlation, as well as  the space-time  distribution of PDC, remain valid even well outside these boundaries, as e.g.  for a 160 fs pulse propagating in  2 mm crystal.  Clearly,  aspects such as the loss of efficiency of the stimulated PDC due to the walk-off are overlooked; these would require more sophisticated descriptions \cite{Gatti2022}, lacking  however the immediacy and simplicity of the model here presented. 
\section*{Methods}
\subsection*{Expansion of phase matching}
We expand the phase matching defined by Eq. \eqref{PM} in Taylor series of the pump variable  $\w_0 = (q_{0x},q_{0y}, \Omega_0)$. 
Provided that the duration and cross section of the pump pulse are large enough  to make  its diffraction and dispersion along the medium negligible
 (and/or the crystal is short enough), quadratic and higher order terms can be neglected, so that:  
\beq
\DD ( \w;   \w_0 - \w)   
- \DD ( \w;   - \w) 
 \simeq \left[ \vec{\nabla} k_{sz}  (-\w)-\vec{ \nabla} k_{pz} (0)    \right] \cdot \w_0
\simeq \left (    k^\prime_s  -k^\prime_p -   k''_s \Omega \right) \Omega_0  - \left( \rho_p  \ex  +\frac{\vec q}{k_s}  \right) \cdot \q_0
\eeq
where in the second passage only the leading terms of the expansion of $ \vec{\nabla} k_{sz}  (-\w)$ around  $\w=0$  (the central frequency and the collinear direction) have been retained, and: 
 $ k^\prime_j= \frac{d k_j}{d\Omega} ({\w=0} )=\frac{1} {v_{gj}} $,   where $v_{gs} $  and $v_{gp} $ denote  the group velocities of the signal and pump wave-packets; 
 $k''_s = \frac{d^2 k_s}{d\Omega^2} (\w=0)$ ; $ \rho_p  \approx \frac{d k_p}{dq_x}$ is the  { \em walk-off} angle of the Poynting vector of the extraordinary pump, assumed  here  in the x-direction.  
The term $ k''_s \Omega \Omega_0$  is usually negligible, unless special points are considered:  for example, with our parameters it would be significant only for  $ |\Omega|$ as large as $   |k'_s-k'_p|/k''_s =  2. 10^{15} s^{-1}$, which is larger than $\omega_s$. 
For a very focused pump, the term $ \frac{\vec q\cdot \q_0}{k_s}  $ may originate the so-called  {\em hot spots} \cite{Perez2015},   but in any case it is relevant only close to a   specific angle $q_x/k = -\rhop $, which we do not consider in our analysis (for our BBO $\rho_p=-3.2^\circ$).  Therefore, under basically  the requirement that diffraction and dispersion of the pump along the medium can be neglected, we do not make a big error by writing
\beq
\DD ( \w;   \w_0 - \w) l_c = \left [ \D(\w)  +    \left ( k^\prime_s  -k^\prime_p \right) \Omega_0  - \rho_p   \ex  \cdot \q_0 \right] l_c = \D(\w)l_c + \taugvm \Omega_0 + \lwoff q_{0x}
 \label{Taylor}
\eeq
\subsection*{Thin crystal solution}
We consider here a particular limit, in which  $g$ is finite, but the crystal is thin enough that  all the  terms of Eq.\eqref{Taylor} are negligible, so that one can set 
\beq
\DD (\w, \w_0-\w) l_c \approx 0 
\label{TC}
\eeq 
for all the modes of interest. 
By  back transforming   the propagation equation \eqref{prop} into direct space, one obtains the  simple parametric equation
\begin{align} 
\frac{\partial  \as }{\partial z}   (\vxi, z )   &=  \frac{g}{l_c}   \, \apc (\vxi ) \, 
 \as^\dagger(\vxi, z)   ,
\label{TC2}
\end{align}
in which space-time points are not coupled, and the parametric gain is modulated by the pump spatio-temporal profile. The solution can be  written as : 
$ \as^{out} (\vxi) = \cosh{ \left[ g |\alpha_p (\vxi) \right]}   \as^{in} (\vxi) + e^{i \phi_p (\vxi)}  \sinh{ \left[ g |\alpha_p (\vxi) \right]}  { \as^{in\, \dagger}} (\vxi) $, which  immediately  gives: 
$\langle  \as^{out} (\vxi)  \as^{out} (\vxi ')  \rangle = \delta(\vxi-\vxi')   
 \frac{1}{2}   \sinh{ \left[2  g |\alpha_p (\vxi)| \right]} e^{i \phi_p (\vxi)} 
$, and 
$\langle  \as^{out\, \dagger} (\vxi)  \as^{out} (\vxi ')  \rangle = \delta(\vxi-\vxi')
  {\sinh^2{ \left[ g |\alpha_p (\vxi)| \right]}}
$. We stress that in principle these are not the second order moments in the  direct space, because  the lowercase operators  $\as $ are connected to the  actual photonic operator $\As $ by the Fourier space transformation \eqref{ajdef}. However, in the  spirit of the thin crystal approximation\eqref{TC},  the propagation phase factors can be neglected,  setting 
$l_c [k_{sz} (\w) + k_{sz}(\w')  ] \approx l_c  k_p $ and  $l_c [k_{sz} (\w) -  k_{sz} (\w') ] \approx 0 $. Therefore, the second order moments in the Fourier space can be obtained by Fourier transforming the above results, as 
\beq
\begin{aligned} 
& \Psi (\w, \w')  \approx e^{ik_p l_c}   \int \frac{d^3 \xi}{(2\pi)^3}   e^{-i  ( \w +\w')   \cdot \vxi}     \, \frac{ 1}{2} \sinh {[ 2 g |\apc (\vxi )| ]}     e^{i \phi_p (\vxi )}   \\
& \Gone(\w, \w')  \approx   \int \frac{d^3 \xi}{(2\pi)^3}   e^{-i  ( \w' -\w)   \cdot \vxi}     \, \sinh^2 {[  g |\apc (\vxi )| ]}   
\end{aligned} 
\label{TC3}
\eeq
These formula coincide with those of the QS model, when the thin crystal limit of  equations \eqref{QSmodel} is taken, because   in the limit \eqref{TC},  $U (\w) \to \cosh{g}$ ,  $ V(\w)  \to \sinh{g}$, and   also the displacement $\vxi_M $  becomes negligible on the pump scale. \par
At this point one might ask where the  precise value  $\vxi_M =  \frac{\lwoff}{2}  \ex + \frac{\taugvm}{2 } \et $  comes from, also considering that it coincides with what observed in the simulations   even for very short pump pulses. Actually, this result was obtained in the context of a slightly more sophisticated model for a thin crystal \cite{Gatti2022}, which for reasons of brevity is not presented here. The  analysis in  \cite{Gatti2022} fully
retains the  effects of  the term   $ \Omega_0 \taugvm + \lwoff q_{0x}$
 in the phase matching
expansion \eqref{Taylor}, and indeed demonstrates that the signal appears at the crystal output displaced by $\vxi_m$ with respect to the pump. 
For reasonably small values of $\Delta= [ (\frac{\taugvm}{\tau_p})^2 + (\frac{\lwoff}{w_p})^2 ]^{1/2}$ , this rigid displacement is basically  the only deviation from  the more naive analysis of  this section,  and for this reason it was incorporated into the QS model. For larger values of $\Delta$,  there is also  \cite{Gatti2022}   the expected  loss of efficiency   due to the lack of overlap  between  the two co-propagating  beams, which instead is neglected by the QS model. 
\subsection*{Numerical simulations}
We considered  the nonlinear propagation equations for the coupled signal and pump operators: 
\bsub
\label{NLeq}
\begin{align} 
\frac{\partial  \as }{\partial z}   (\w_s, z )   &=  \chi  \int 
 \frac{d^3 \w_p }  {(2\pi)^{\frac{3}{2}} }  \ah_p(\w_p ,z) 
 \as^\dagger  (\w_p -\w_s, z)    e^{-i \DD (\w_s, \w_p-\w_s) z }  
\label{NLs} \\
\frac{\partial  \ah_p }{\partial z}   (\w_p, z )   &= -\frac{ \chi}{2}  \int 
 \frac{d^3 \w_s }  {(2\pi)^{\frac{3}{2}} }  \as (\w_s  ,z) 
 \as (\w_p -\w_s, z)  e^{i \DD (\w_s, \w_p -\w_s) z }  
\label{NLp}
\end{align} 
\esub
Simulations of these equations were performed in the framework of the quantum to classical correspondence, in the Wigner representation, which provides the symmetrically ordered moments of observables. More precisely, we used a truncated Wigner representation \cite{Werner1995,Gatti1997},  in which the quantum operators  are replaced  by c-number fields that evolve with equations  formally identical to Eqs.\eqref{NLeq},  and  the  quantum noise contributes only  through  the vacuum input fluctuations.   In our simulations, they are modeled by taking
 Gaussian  white noise as initial condition for the signal. 
 The input pump, which is assumed to be  a high-intensity coherent pulse, is modeled  by  a Gaussian profile in space and time.  For all the considered simulation parameters, we found that the  pump was nearly undepleted. 
\par
We considered  PDC in a 2mm BBO crystal,   tuned for type I e-oo collinear phase-matching at degeneracy  when pumped by a laser at $\lambda_p=515 $nm (angle  of propagation $  \sim 23.29^\circ$ with the optical axis).  In these conditions 
 the signal spectrum exhibits the  characteristic X-shape   shown in Fig.\ref{fig_Xspectrum}d,e. 
Numerical integration is performed through a second-order pseudo-spectral (split-step) method \cite{numrec}. Temporal dispersion 
 and  diffraction  are taken into account at any order  by using the complete Sellmeier relations for the  refractive index of the material \cite{Eimerl87}. 
Since our results need averages performed over a   large number of independent realizations (up to $10^5$ for small $g$)  , which are extremely time-consuming, we performed 2D+1 simulations, restricting to one transverse spatial dimension, i.e. the one along the walk-off direction. 
Our numerical grid is 512$\times$512 pixels in the $q_x$ and $\Omega$ directions, spanning  the symmetric bandwidths $-0.3{\rm \mu m}^{-1}<q_x<0.3{\rm\mu m}^{-1}$, and $-3.84\times 10^{14}{\rm s}^{-1}<\Omega<3.84\times 10^{14}{\rm s}^{-1}$  ($850{\rm nm} <\lambda<1304$nm)  around the collinear direction and the degenerate frequency (see Fig. \ref{fig_Xspectrum}).  In the $q_y$ direction we took a single pixel, which may simulate  a narrow slit in the far-field of the source that selects  a single mode around $q_y=0 $, as e.g. done for frequency-resolved detection  by  means of an imaging spectrometer \cite{Allevi2014}. 
\par
The biphoton correlation and the coherence function defined by Eq.(\ref{Psi}) and Eq.(\ref{G1}) were evaluated by performing ensemble averages over a  large number  (from $10^3$ to $10^5$, depending on the gain) of  independent realizations, obtained by integrating    the propagation equations (\ref{NLeq}) starting from independently and randomly generated initial conditions.  In order to boost  the convergence rate of the simulations  we  also exploited the translation  invariance of the field statistics in the central region of the Fourier plane where the spectrum  is nearly uniform. We  considered  
 the  rectangular region  $R=[-0.1 {\rm \mu m}^{-1},0.1 {\rm \mu m}^{-1}]\times [0,1.2\, 10^{14} {\rm s}^{-1}]$,
containing 
$M=13338 
$   pixels
and in each realization  calculated the discrete convolutions $C_{corr}(\w_k)= \frac{1}{M}\sum_{\w_j} A_s^{out}(\w_j  )A_s^{out}(-\w_j +\w_k)$
and $ C_{coh}(\w_k)= \frac{1}{M}\sum_{\w_j} A_s^{\star\, out}(w_j  )A_s^{out}(\w_j +\w_k)$, where the  $ \w_j$ run over the  pixel coordinates inside R.
 Examples of the biphoton correlation obtained in this way  are shown in  Fig.\ref{fig_corr600} and Fig.\ref{fig_corr160}.  
The standard deviations  $\Delta \Omega$ and $\Delta q$  here  reported  were evaluated  by using the $\left|\Psi\right| $ and $ | \Gone |$  thereby obtained (after correcting the $\Gone$ in order to pass from symmetric to normal ordering).  
A delicate point is represented by the residual noise of the Wigner simulation, still visible e.g. in Fig. \ref{fig_corr160}a, that may artificially enhance  $\Delta \Omega$ and $\Delta q$: for this reason the variances along each Fourier coordinate  were calculated over a reduced region, covering 8 times the standard deviations \eqref{varcorr}  of the QS model.  This number seemed  a good compromise between the need to cover the entire peak and to avoid including too much residual noise.
\section*{Author contributions statement}
A.G. conceived the QS model and studied its predictions, E.B. wrote the code and conducted the numerical simulations. All the Authors collaborated to the statistical analysis and discussed the results.

The datasets used and analysed during the current study are available from the corresponding author on reasonable request. 
%
The authors declare no competing interests.


\end{document}